# Design and Production of Specifically and with High Affinity Reacting Peptides (SHARP®-s)

by

*Jan C Biro*

HOMULUS FOUNDATION, 612 S. Flower Str., #1220, 90017 CA, USA
jan.biro@att.net
www.janbiro.com

## Abstract

### Background

A partially random *target* selection method was developed to design and produce affinity reagents (*target*) to any protein *query*. It is based on the recent concept of *Proteomic Code* (for review see Biro, 2007 [1]) which suggests that significant number of amino acids in specifically interacting proteins are coded by partially complementary codons. It means that the 1st and 3rd residues of codons coding many co-locating amino acids are complementary but the 2nd may but not necessarily complementary: like 5'-AXG-3'/3'-CXT-5' codon pair, where X is any nucleotide.

### Results

A mixture of 45 residue long, reverse, partially complementary oligonucleotide sequences (*target* pool) were synthesized to selected epitopes of *query* mRNA sequences. The 2nd codon residues were randomized. The *target* oligonucleotide pool was inserted into vectors, expressed and the protein products were screened for affinity to the *query* in Bacterial Two-Hybrid System. The best clones were used for larger-scale protein syntheses and characterization. It was possible to design and produce specific and with high affinity reacting (Kd: ~100 nM) oligopeptide reagents to GAL4 *query* oligopeptides.

### Conclusions

Second codon residue randomization is a promising method to design and produce affinity peptides to any protein sequences. The method has the potential to be a rapid, inexpensive, high throughput, *non-immunoglobulin* based alternative to recent *in vivo* antibody generating procedures.

## Introduction

According to the Human Genome Project, the estimate for the total number of genes in the human genome has been revised down from 30,000 – 35,000 to 20,000 – 25,000 [2].

Despite the fairly low number of genes, the best estimates for the total number of proteins encoded by the human genome (the proteome) remain anywhere from 300,000 to 1 million. Ongoing efforts to study the proteome have kept the research antibody industry flourishing. Also reaping financial benefits are motivating companies that provide reagents and devices required for antibody-related protocols. With basic science and drug developing researchers creating a tide of demand, the revenue stream from antibody-led protein hunts won't be drying up anytime soon [3].

Revenue from antibodies for therapeutic and diagnostic uses is expected to grow at an average annual growth rate of 11.5%, according to a 2005 report, "Dynamic Antibody Industry," published by the Business Communications Company [4]. With an estimated market of $15 billion in 2005, revenues should reach $26 billion by 2010.

The original procedure to induce antibodies (Ab) is the *in vivo* immunization. The success of this polyclonal antibody production in animals is variable, often not predictable, takes several weeks, require the use of well purified antigens [5]. Monoclonal antibody production uses animals only as the source of biological machinery which is responsible for the immune response (B lymphocytes). The Ab production itself takes place *in vitro* [6]. It makes the monoclonal Ab production cheaper, faster and more reliable. In fact, monoclonal antibodies produced by animal immunization remain the 'gold standard' of affinity reagents. They are relatively renewable, can usually be made with high specificity and affinity for their target and can be used in common biochemical assays such as Western blotting [7], ELISA [8] and different branches of immunochemistry [9]. But the traditional monoclonal antibody has its drawbacks. Its production can be challenging, time-consuming and costly.

Additional concerns associated with Ab-s which are seriously limiting their therapeutic applications are their size and their origin. Antibodies are large and complex proteins (>150 KDaltons) which are much larger than necessary for antigen recognition and binding. They are antigenic themselves and are carrying the species characteristics of their origin. Size reduction, like single-chain variable (scFv) antibody fragments [10] and "humanization" [11] help to solve these problems.

A moderate discomfort, even if not a ban, of Ab production is that we still not know exactly how antibodies are made by the immune system and therefore it is not possible to reproduce it *de novo*, without "borrowing" the technique of a living organism.

So there is a lot of interest in identifying novel affinity reagents that would be less expensive and quicker to produce.

*Affibodies* [12] were among the first non-immunoglobulin-based affinity reagents. These small molecules are based on a bacterial receptor

(Staphylococcus aureus protein A), and use combinatorial protein engineering to introduce random mutations in the affinity region [13]. Protein Z is a 58-residue three-helix bundle domain derived from staphylococcal protein A (SPA), which binds to the Fc portion of IgG from different species. By simultaneously randomizing 13 amino acid positions located at the two helices making up the Fc-binding face of protein Z, binding proteins (*affibodies*) capable of binding to desired targets have been selected by using phage display technology.

Another non-immunoglobulin-based affinity reagent that is becoming more widely used is the *aptamer*. Made of DNA, RNA or modified nucleic acids and typically 15–40 bases in length, aptamers have a stable tertiary structure that permits protein binding through van der Waals forces, hydrogen bonding and electrostatic interactions. Early studies showed that aptamers can be highly specific for target proteins, with the ability to distinguish between related members of a protein family [14].

No one of the above mentioned methods is able to satisfy the emerging need to produce these reagents at a truly *high-throughput* scale. The estimated number of affinity reagents (antibodies or else) needed to monitor the human proteome is probably not less than the number of different proteins (including splicing variations, but not the unknown number of folding variations), which is around 0.3-1.0 millions.

A completely different approach to understand and utilize the nature of specific protein-protein interactions is the concept of the *Proteomic Code*. The Proteomic Code is a set of rules by which information in genetic material is transferred into the physico-chemical properties of amino acids. It determines how individual amino acids interact with each other during folding and in specific protein–protein interactions. The Proteomic Code is part of the redundant Genetic Code [1]. This, 25-years-old theory states, that significant number of amino acids co-located on the specifically interacting protein-protein interfaces are coded by complementary codons.

The original concept expected perfect complementarity coding of co-locating amino acids. There are hundreds of experiments from reliable laboratories which are suggesting the validity of this expectation (for review see [1, 15]). However there are many (to many) exceptions.

Recently, bioinformatical studies suggested that co-locating amino acids are coded by partially complementary codons, where the 1$^{st}$ and 3$^{rd}$ codon residues are complementary in reverse orientation, but the 2$^{nd}$ codon residue is not necessarily complementary. This second generation *Proteomic Code* may be described by the ***3'-NXN-5'/5'-nXn -3'*** formula, where **N** and **n** denote complementary base pairs, while **X** indicates any nucleotide. A method (The Method in this text) to *Design and Production of Specifically and with High Affinity Reacting Peptides (SHARP®-s)* is built on this formula [16].

### *General aspects of the Method*

The protein which is used to generate affinity peptides is called ***query (Q)*** in this description and it is analog to the terms *ligand, antigen or bait*. Query is one protein sequence that the target protein, designed and produced with the Method, will specifically interact with.

The generated affinity peptide is called ***target (T)***, and it is analog to the names *receptor, antibody or hit*. Target proteins are protein sequences which are designed by the Method to specifically interact with the query protein sequence.

The query and target are expected to react with each other specifically (able to distinguish between related but not identical peptide sequences) and with high affinity (Kd is at least in microM range).

***Target Oligo-Nucleotide Pool (TONP)*** is designed by using a ***Target Ologo-nucleotide Template (TONT)*** which is a nucleic acid sequence containing 2/3 defined and 1/3 undefined (any) nucleotides (X). A TONT, which contains 15 undefined nucleic acid residues, (defining a pool of $4^{15}=10^{9}$ different oligo-nucleotides, TONP) will be translated into the corresponding number of oligopeptides.

Expression of TONT will result in the syntheses of a large number of different oligo-peptides, called Target Oligo-Peptide Pool (TOPP). Those oligopeptides which satisfy the criteria for specific, high affinity reactions with the query protein are called SHARP®-s.

The flow of SHARP® production may comprise the following main steps (Figure 1).

**Figure 1**

**Design and Production of Specifically Interacting Proteins**

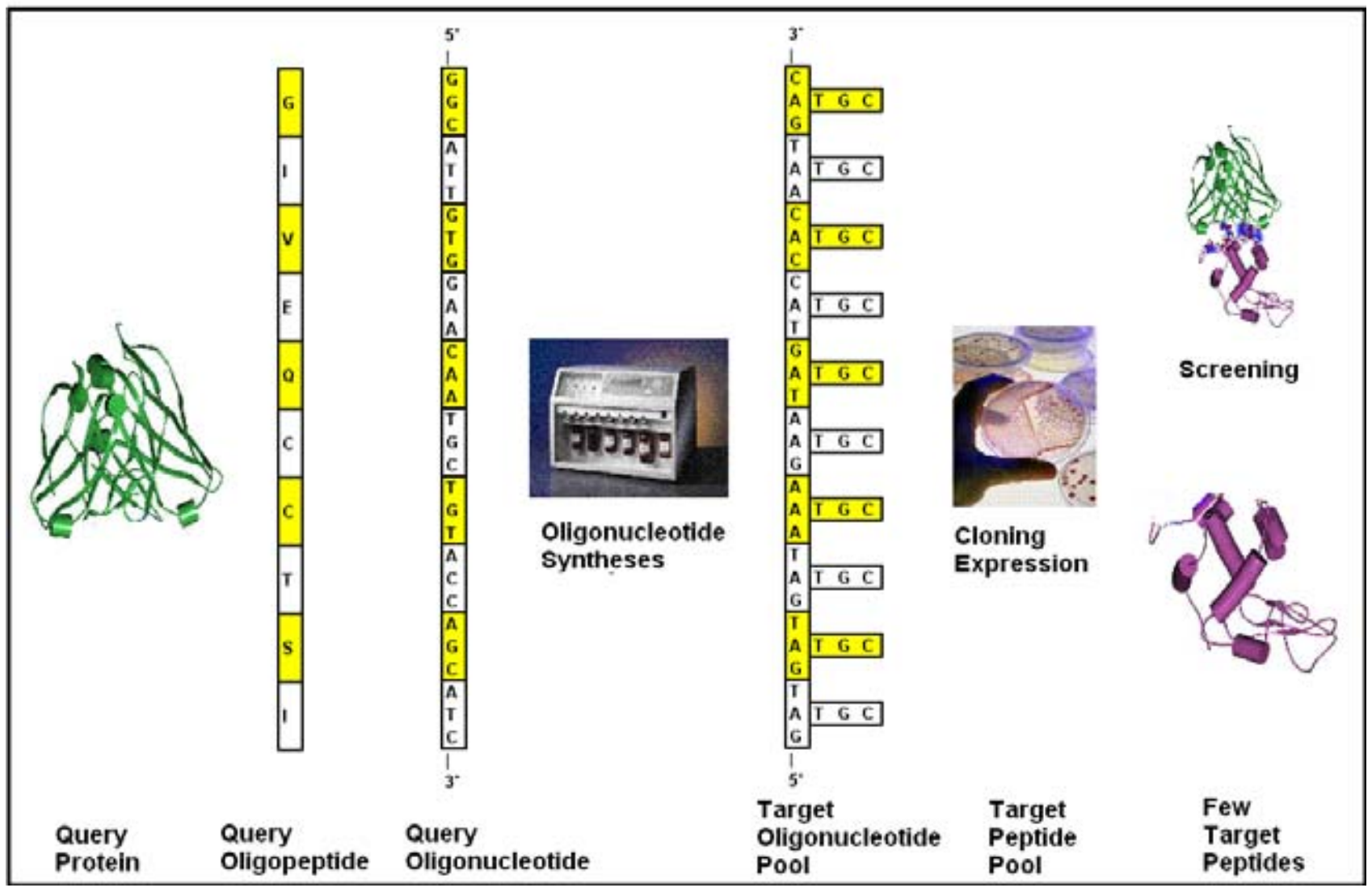

**Figure 1: Schematic illustration of design and production of Specifically interacting proteins SHARP®s**

**1. Query selection**

- The length of query peptide is limited to about 15 amino acids, depending on the available screening method (see below). Expression library utilizing phages [17] may contain maximum $10^9$ different clones (which is ~ $4^{15}$) while the recent maximal capacity of yeast based libraries [18] is much less, about $10^7$ (which is ~ $4^{12}$). This limits the number of variable bases to 15 and 12, respectively.
- Query may be part of a much larger protein. In these cases physicochemical, structural considerations might guide to select the most “promising” query sequences.
- The concept of selecting a particular sequence of a protein as a target to selected complementary peptide or vice- versa is a formidable challenge. Epitopes exist as continuous or discontinuous and as such are complex when antibodies are raised against them. To have an alternative system where protein-protein or protein- peptide interaction are involved the spatial conformation have to be taken into account.
- It is necessary to know the correct coding sequence of the query. A nucleic acid sequence derived by back-translation, using any Codon Usage Frequency Table, is not satisfactory.

**2., Design the target CDS**

- The CDS of target oligopeptides (TONP) is derived from the CDS of the query, using the 3’***-NXN-5’/5’-nXn-3’*** formula [1] or

***3’-_NNN_NNN_NNN_-5’ .....QUERY***
***5’-_nXn_nXn_nXn_-3’ .....TARGET OLIGO-NUCLEOTIDE TEMPLAT (TONT)***

N and n are complementary bases, and X is any nucleotide.

- X should always be the central codon residue, therefore knowing the correct translation frame is essential. Frame-shifts are not permitted.
- The limitation of CDS length is around 45 nucleotides, depending on the capacity of the expression system.

**3., Additions to the target CDS**

The target CDSs (TONP) will be inserted into vectors for expression, therefore sequence additions are necessary

- Add restriction enzyme sites at both ends (choose restrictions enzymes which have no cut site in the designed target sequences);
- Add known sequences (after the start signal) which will identify the expressed mRNA and proteins and able to confirm the correct translation frame, if necessary. A short polypeptide marker sequence useful for recombinant protein identification and purification [19].

**4., Synthesize the target CDSs (TONP).**

There are 3 different nucleic acid synthesizing methods, which might be suitable for the synthesis of TONP:

a., *Chemical syntheses of oligonucleotides using synthesizers.* A mixture containing equal concentration of A, T, G, C residues should be used at the X residues [20].

b., *Site directed mutagenesis by Error-Prone PCR*.
This method is based on the incorporation of mutagenic dNTP analogs, such as 8-oxo-dGTP and dPTP, into an amplified DNA fragment at the X positions by PCR. The mutagenic dNTPs are eliminated by a second PCR step in the presence of the four natural dNTPs only, resulting in a rate of mutagenesis of up to 20%. (Mutagenesis Kits are commercially available) [21].

c., *Codon-varied oligonucleotide synthesis for synthetic shuffling*. This method uses codons instead of single nucleotides for CDS synthesis [22]. This method can make it easier to position the mutations (X) always into the central codon residue.

The result of these syntheses will be a mixture of more or less different oligonucleotides which differ from each other at one ore more central codon residue, but they are identical at the 1$^{st}$ and 3$^{rd}$ codon residues. The number of expected target oligonucleotide sequences is $4^c$, where c is the number of codons (amino acids) in the designed target sequences. This mixture of different target sequences is called the Target Oligonucleotide Template Pool (TONP).

It is necessary to validate the correctness of nucleic acid synthesis to make sure that the sequences in TONP contain the TONT pattern.

**5., Insert the TONP into cloning vectors** using restriction enzymes and ligase [23]. Commercially available vectors contain numerous restriction enzyme cut sites where the TONP sequences (containing the properly prepared 5' and 3' termini) should be inserted with ligase reaction. It is important to pay attention to the orientation and translational phase of the inserts.

**6., Insert** the TONP containing vectors into the chosen **screening system** (phage, yeast or bacteria) [23].

**7., Test** that the expression of TONP works properly, i.e.
- TONP mRNAs are present;
- the TONP mRNAs contain the TONT pattern (correct translation frame is used, orientation is correct);
- Detect the signal protein (if it's CDS was added to the TONT). This is a further indication that translation is correct, no frame shift occurs and the target oligo-peptides (TOPP) are correctly expressed.

**8., Prepare the query protein.**

Some screening systems (two-hybrid yeast or bacterial systems) request the expression of the query too. In these cases prepare the query accordingly to the systems requirements and test the correctness of query expression.

**9. Expression/cloning library construction**

The TOTP might be expressed and multiplied in yeast [24], bacteria (using plasmid or phage vectors) [25], which are providing the TOPP expression library. This libraries are expected to contain and express about $10^6$-$10^9$, more or less different, oligopeptides. Target Oligopeptide Library is a partially (33%) random library, because the central residues of the codons are randomly selected.

**10. Screening of the TOPP expression library**

There are at least two different methods which are suitable for screening the clones

a. protein-*fragment complementation assays (PCA)* [24]

This assay is utilized by the bacterial and yeast two hybrid methods (Figure 2). In these types of assays an active enzyme is dissected into inactive fragments and the fragments are fused to the test proteins. Interaction between the test proteins brings the inactive enzyme fractions together and restores the original function of the enzyme. This enzyme function is than detected by a simple assay (colorimetry, fluorimetry, colony survival).

b. *phage display technique* (for review see [25], Figure 3).

In these assays one of the test proteins (target) is fused to phage proteins and expressed on the surface of phages. The other test protein (query) is attached to some solid surface. Interaction between test proteins will attach the phase to the solid surface and select the phages which are producing the interacting test protein.

**Figure 2**

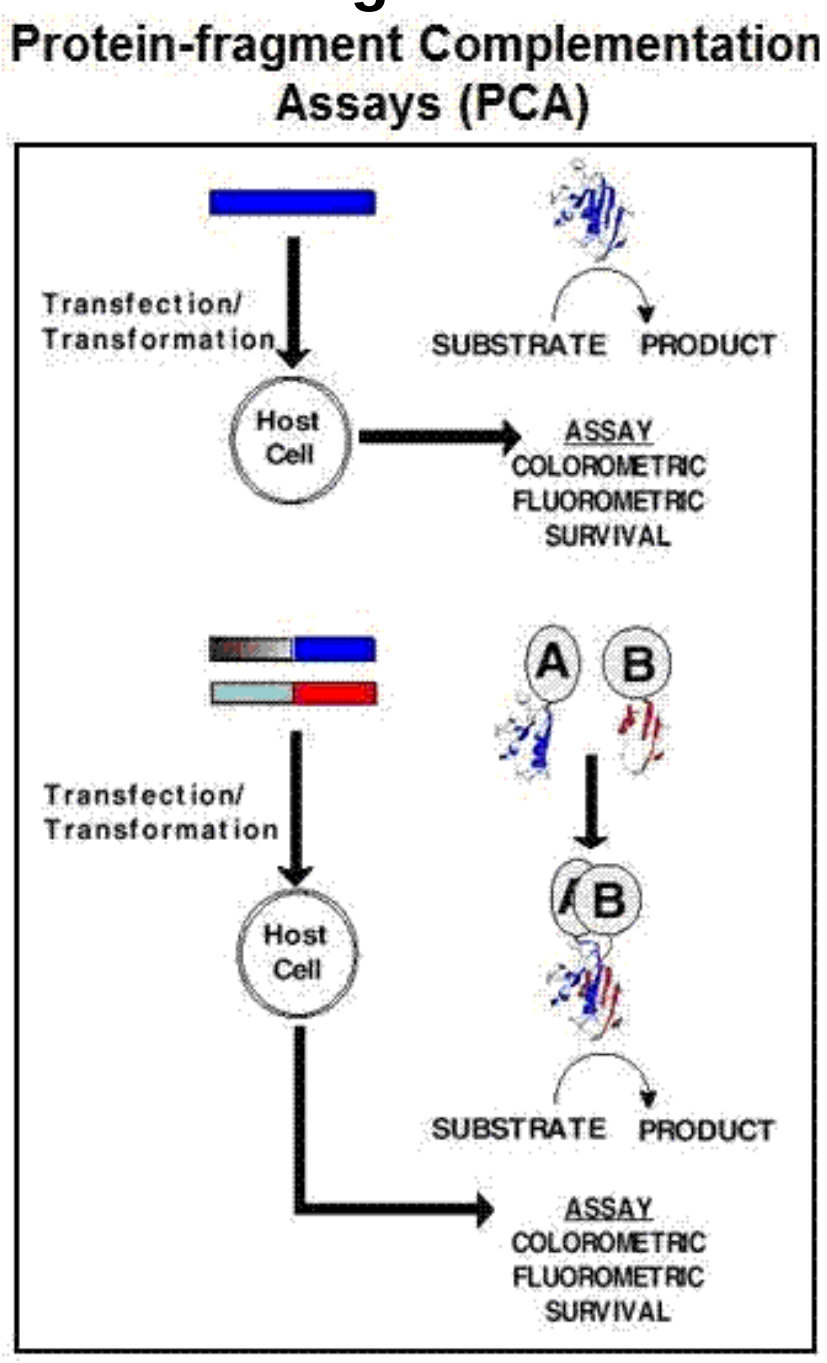

**Figure 2: General description of a PCA** [26]. The gene for a protein or enzyme is rationally dissected into two or more fragments. Using molecular biology techniques, the chosen fragments are sub cloned, and to the ends of each, proteins that either are known or thought to interact are fused. Co-transfection or transformation of these DNA constructs into cells is then carried out. Reassembly of the probe protein or enzyme from its fragments is catalyzed by the binding of the test proteins to each other, and reconstitution is observed with some assay.

**Figure 3**

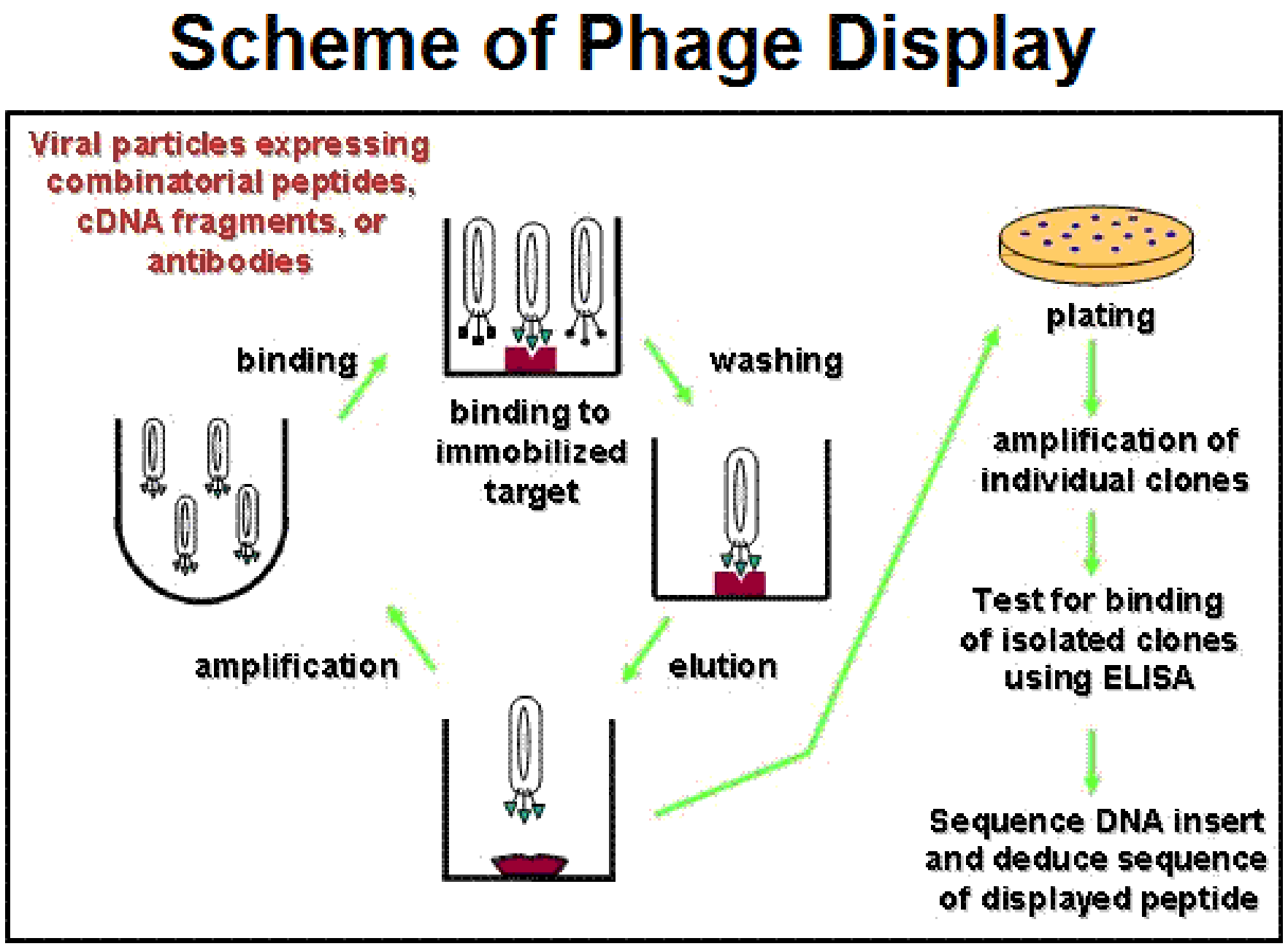

**Figure 3: Scheme of phage display** [27].

11. **Select** the best clones and continue the cloning until monoclonal stadium. Save the best clones. The “best” clones are those which are displaying the largest number of target proteins with highest affinity to the query protein. Numerous methods are available for manual or automated library handling and screening of surface displayed interacting (binding) gene products [28-31].

12., The best clones should be further propagated to obtain larger quantity of the desired SHARP®-s for purification and **testing for physicochemical properties** (binding specificity and Kd).
- Large number of methods is known for Kd determination. Label-free real-time interaction analyses **using surface plasmon resonance and a sensor chip technology**, is one of the latest technologies [32, 33]. These methods work on crude bacterial extracts, target protein purification is not necessary.

13. Extract and **re-sequence the target CDSs** producing the best targets.

14. **Re-sequence the best target oligo-peptides and their mRNA**. This step is not absolutely necessary. Finding the proper SHARP® might be satisfactory. However re-sequencing is the source of important information for further experiments. For example

- re-sequencing is the final confirmation, that the target peptides are found by design and not only by chance.
- re-sequencing and analyzing the query/target sequences might be the source of valuable knowledge about the further rules of protein-protein interactions.
Therefore re-sequencing is highly recommended.

15., **Visualization of query/target complex** (NMR) may be an additional step to gain further knowledge about the protein folding and interactions. NMR analyses for protein-protein interactions in solution are very important methods in understanding specific, macromolecular interactions [34-35]. However the huge complexity of NMR investigation techniques shouldn't be underestimated.

16., **Iterate** the procedure 1-15 using another query.

## Example for application of the Method

*BacterioMatch II two-hybrid system*

The system chosen was the BacterioMatch II two-hybrid system commercially available from Stratagene [36]. This system utilizes a *bait vector* (pBT) and a *target vector* (pTRG). The bait vector contains a portion of the λ cI gene and the target vector contains a portion of the alpha subunit of RNA polymerase. The bait and target peptides (or proteins) are genetically fused to the cI gene in pBT and the RNA polymerase subunit in pTRG, respectively. Inside a cotransfected bacterial cell, the λ cI gene product binds to an λ operator on a reporter plasmid. The bait portion of the construct is available to interact with the target portion of the target vector. When an interaction occurs (the bait and target bind to each other), the RNA polymerase subunit is in close proximity to a weak RNA polymerase binding site (from the *lac*Z promoter) on the reporter plasmid. This binding allows the RNA polymerase to transcribe a pair of reporter genes, HIS3 and aadA. The HIS3 gene allows the bacterial cells to grow on medium lacking histidine (or more accurately containing the histidine antagonist 3-amino-1, 2, 4-triazole or 3AT). If the binding interaction is strong enough, the aadA gene is also transcribed conferring streptomycin resistance on the cells. Double selection on 3AT and streptomycin containing plates reduces the number of false positives. The BacterioMatch II kit is supplied with a pair of control vectors containing portions of the yeast *GAL*4 and *GAL*11 genes. The interaction of these gene products is both highly specific and tight. Cells from a transfection using both of these controls produce many colonies on doubly selective plates. (Figure 4).

**Figure 4**

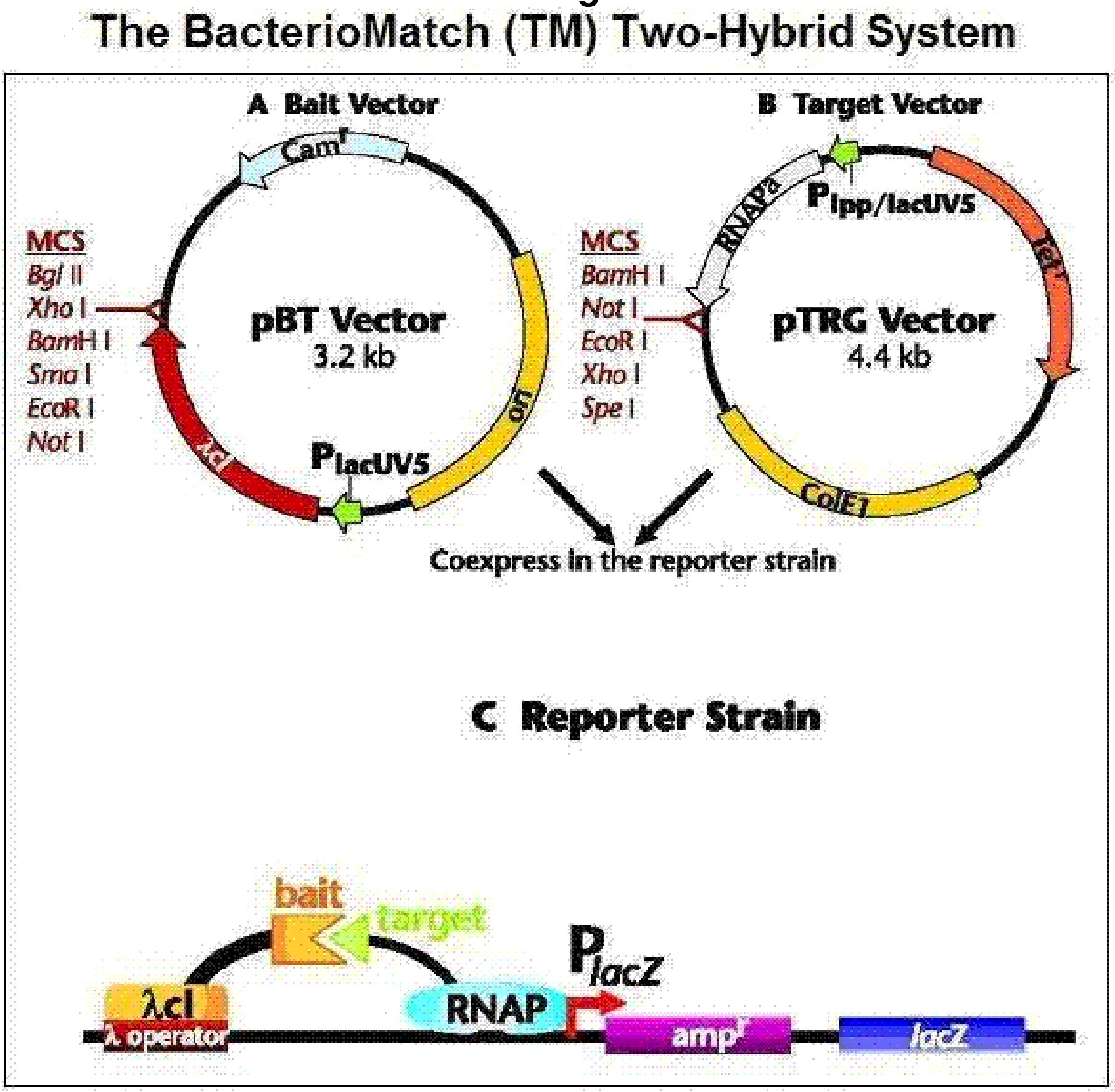

**Figure 4: The BacterioMatch™. Two-hybrid system** [36]

*Outline of Experimental Design*

We designed a series of target peptide gene sequences that could potentially bind to the control vector pBT-LGF2 (containing the λ cI gene). These target oligonucleotides (the Target Oligo Nucleotide Pool) were synthesized by Retrogen [37], digested with appropriate restriction enzymes, and ligated into similarly digested pTRG. The pBT-LGF2 and pTRG-TONP were cotransfected into E. coli.

*Materials and Methods.*

Analytical grade chemicals were used.

The sequence of the GAL4 portion of the pBT-GAL4 obtained from the Strategene website was inconsistent with the size of the GAL4 protein as described in the Strategene literature. Forward and reverse oligonucleotide primers for sequencing pBT were synthesized (Retrogen) and the nucleotide sequence of the GAL4 protein was determined. This sequence was then matched by to the published sequence to establish the exact portion of the GAL4 protein contained in the pBT-LGF2 vector.

TONTs were designed by to match the protein sequence predicted from the newly determined nucleic acid sequence of GAL4. The sequences in TONP included the restriction sites for *Bam*HI and *Not*I (Figure 5). One strand of the sequences in TONPs was chemically synthesized using multiple bases at various positions (denoted with an X in Fig. 5). The second strand was produced by a primer extension reaction; the products of this reaction were analyzed by gel electrophoresis (20% PAGEgel, 0.25*M* Tris acetate buffer at pH 7.6). After the primer extension reaction, the TONP were desalted (Sephadex G-50 Micro Columns), digested with BamHI and NotI, heated to inactivate the restriction enzymes, and desalted again.

**Figure 5**

**Design of SHARP®s**

## A. Query Selection

```
>1HBW:A REGULATORY PROTEIN GAL4
1       9       17     23       31                                 57
.       .       .      .        .                                  .
TRAHLTEVESRLERLEQLFLLIFPREDLDMILKMDSLRDIEALLTGLFVQDNVNKDA
     HHHHHHHHHHHHH SSSSSSS   HHHHHHTTHHHHHHHHHH  S SS

>K01486_SCGAL4_DIMDOM | ND, 171 bases, 1D18 checksum.
5'-- ACTAGGGCACATCTGACAGAAGTGGAATCAAGGCTAGAAAGACTGGAACAGCTATTTCTACTGATTTTTCCTCGAGAAGACCTTGACATGATTTTG
H2N-  T  R  A  H  L  T  E  V  E  S  R  L  E  R  L  E  Q  L  F  L  L  I  F  P  R  E  D  L  D  M  I  L
```

```
>Query 1_ ESRLERLEQLFLLIF (GAL4 09-23AA)
5'--GAATCAAGGCTAGAAAGACTGGAACAGCTATTTCTACTGATTTTT
H2N- E  S  R  L  E  R  L  E  Q  L  F  L  L  I  F

>Query 2_QLFLLIFPREDLDMI (GAL4 17-31AA)
5'--CAGCTATTTCTACTGATTTTTCCTCGAGAAGACCTTGACATGATT
H2N- Q  L  F  L  L  I  F  P  R  E  D  L  D  M  I
```

## B. Target Oligo-Nucleotide Template (TONT) Design

```
>TARGET TEMPLATE to Query 1_ESRLERLEQLFLLIF (GAL4 09-23AA)-ssDNA-sense
5'-GGATCC-[AxAAxTCxGTxGAxATxGCxGTxCCxGTxTTxCTxGCxTTxATxC]-CGGCCGC-3'
|--BamHI--[    VARIABLE SEQUENCE, TONT  45 NA           ]-NotI----|

>TARGET TEMPLATE to Query 2_QLFLLIFPREDLDMI (GAL4 17-31AA)-ssDNA-sense
5'-GGATCC-[AxTCxTGxCAxGGxCTxCTxGAxGAxAAxTCxGTxGAxATxGCxG]-CGGCCGC-3'
|--BamHI--[    VARIABLE SEQUENCE, TONT  45 NA           ]-NotI----|
```

**Figure 5: Design of SHARP®s.**

Two 15 amino acid (AA) long oligopeptides (Query 1 & 2) were selected from the GAL4 sequence (underlined). Helices (H) and sheets (S) are indicated below the primary GAL4 sequence. Coding sequences of these queries were used to design TONT. Codons are emphasized by yellow boxes. Restriction enzyme cut sites are indicated by blue boxes.

pTRG was digested with *Bam*HI and *Not*I. Approximately 160 ng digested vector (1 µL), 6 to 10 µL digested TONP, and 10 X ligase buffer, ATP, and water were placed in a final volume of 19 µL along with 1 µL of T4 DNA ligase. The reaction mix was incubated overnight at 4° C. XL1-Blue MRF’ Kan cells (Strategene) were transfected with the ligation mixes and plated on non-selective medium containing kanamycin and tetracycline. Several colonies were picked and plasmid DNA was prepared using Wizard Plus SV Miniprep kits [38]. Plasmid DNA was digested with *Xba*I and the digestion products separated on a 1 % agarose gel to confirm the insertion of TONP into the vectors. An XbaI digestion should result in the release of a fragment approximately 845 base pairs in length if there is no insert or 922 base pairs with the TONP inserted.

BacterioMatch II electro competent reporter cells were cotransfected with 50 ng pBT-LGF2 and either 6 or 10 µL of the pTRG-TONP ligations. The cotransfected bacteria were plated on M9 minimal His-dropout medium, M9 minimal His-dropout medium containing 3-AT, and M9 minimal His-dropout medium containing 3-AT and streptomycin media as presented in the BacterioMatch II product insert. IPTG was added to induce expression of the bait and TONP derived fusion proteins, TOPP.

Appropriately sized TONP sequences were synthesized and ligated into the pTRG vector supplied with the BacterioMatch II Two-Hybrid System Kit. Diagnostic restriction enzyme digestions indicate that an appropriately sized DNA oligonucleotide was present in 3 of 4 colonies tested. A control cotransfection with pBT-LGF2 and pTRG-GAL11P resulted in the expected number of colonies under all of the conditions tested.

Some colonies that grew on doubly selective medium (containing 3-AT and streptomycin) were selected for sequencing. The colonies were picked using a pipette tip, placed in 10 mL LB containing kanamycin (100 mg/L), chloramphenicol (5mg/L) , and tetracycline (10 mg/L) and incubated overnight at 30° C. The entire 10 mL culture was processed for plasmid DNA using a Wizard Plus SV Miniprep kit. The Miniprep DNAs were sent to Retrogen for sequencing. Forward and reverse oligonucleotide primers for sequencing pTRG were synthesized (Retrogen) and the nucleotide sequences of the three inserts were determined (Figure 6).

**Figure 6**
**Query and Target Sequences**

```
HOOC---- L  I  M  D  L  D  E  R  P  F  I  L  L  F  L  Q  E  L  R  E  L  R  S  E  V  E  T  L  H  A  R  T     - Query Protein
3’------GTTTTAGTACAGTTCCAGAAGAGCTCCTTTTTAGTCATCTTTATCGACAAGGTCAGAAAGATCGGAACTAAGGTGAAGACAGTCTACACGGGATCA-5’ - Query CDS
5’------------------------GGATCC-[AxAAxTCxGTxGAxATxGCxGTxCCxGTxTTxCTxGCxTTxATxC]-CGGCCGC----------------3’ - TONT 1
5’------------------------GGATCC-[ATAATTCTGTTGATATTGCTGTCCCTGTTTTCCTTGCTTTTATCC]-CGGCCGC----------------3’ - Target Oligo 1.1
5’------------------------GGATCC-[ATAATTCTGTGGATATTGCTGTCCCTGTTTTCCTGGCTTTCATCC]-CGGCCGC----------------3’ - Target Oligo 1.2
5’------------------------GGATCC-[ATAATTCTGTTGATATTGCGGTCCCTGTATTACTTGCGTTTATAC]-CGGCCGC----------------3’ - Target Oligo 1.3
5’-GGATCC-[AxTCxTGxCAxGGxCTxCTxGAxGAxAAxTCxGTxGAxATxGCxG]-CGGCCGC---------------------------------------3’ - TONT 2
5’-GGATCC-[ATTCTTGACAGGGAGTCCTTGAGGATAATCCTGTTGATATTGCGG]-CGGCCGC---------------------------------------3’ - Target Oligo 2.1
5’-GGATCC-[ATTCTTGTCAGGGTGTCCTTGAGGAGAATTCTGTTGATATCGCCG]-CGGCCGC---------------------------------------3’ - Target Oligo 2.2
5’-GGATCC-[ATTCTTGGCACGGAGTACTCGACGAGAATTCTGTTGAGATCGCTG]-CGGCCGC---------------------------------------3’ - Target Oligo 2.3

H2N-------------------------------- I  I  L  L  I  L  L  S  L  F  S  L  L  L  S  ---------------------COOH  - Target Peptide 1.1
H2N-------------------------------- I  I  L  W  I  L  L  S  L  F  S  W  L  S  S  ---------------------COOH  - Target Peptide 1.2
H2N-------------------------------- I  I  L  L  I  L  R  S  L  Y  Y  L  R  L  Y  ---------------------COOH  - Target Peptide 1.3
H2N-------  I  L  D  R  E  S  L  R  I  I  L  L  I  L  R  ----------------------------------------------COOH  - Target Peptide 2.1
H2N-------  I  L  V  R  V  S  L  R  R  I  L  L  I  S  P  ----------------------------------------------COOH  - Target Peptide 2.2
H2N-------  I  L  G  T  E  Y  S  T  R  I  L  L  R  S  L  ----------------------------------------------COOH  - Target Peptide 2.3
```

**Figure 6: Query and Target Sequences**

Target 1 & 2 sequences were designed and produced to interact with Query 1 & 2 (in Figure 5). Multiple sequence aligned (MSA) Nucleic acid and protein oligos are indicated. Codons are indicated by boxes (yellow) in the nucleic acid sequences as well as restriction enzyme cut sites (blue). Most common amino acids are also indicated (yellow background) in the protein sequences.

The best clones were further propagated, extracted and the crude bacterial extracts were used for preliminary characterization of binding properties to pure Gal4 protein. This was performed using surface plasmon resonance and a sensor chip technology [32]. The Gal4 query binding to the selected targets is specific and Kds are ~ 100nM (these results are not shown).

## Discussion and Conclusions

SHARP® design and production is a novel method to obtain affinity peptide reagents. It was derived from the concept of *Proteomic Code*. Methods based on this original concept of the Proteomic Code (perfect complementary coding of interacting peptides) were developed in many laboratories and were remarkably successful in many (but not all) cases. These methods never became widely accepted because of the poor predictability of the results [1].

However new bioinformatics tools [39] and a large protein structure database (PDB) became available. Statistical analyzes of amino-acid co-locations in real protein structures revealed that the redundant Genetic Code contains even information necessary to protein folding and interactions [40, 41].

Recent observations suggest, that naturally co-locating amino acids (in- and between specifically interacting proteins) are preferentially coded by partially complementary codons. The Method above is to replicate this biological phenomenon and try to design specifically interacting oligo-peptides. It is assumed, that protein-protein interactions are determined already on the amino acid level (Figure 7).

**Figure 7**
**Forms of Peptide to Peptide Interactions**

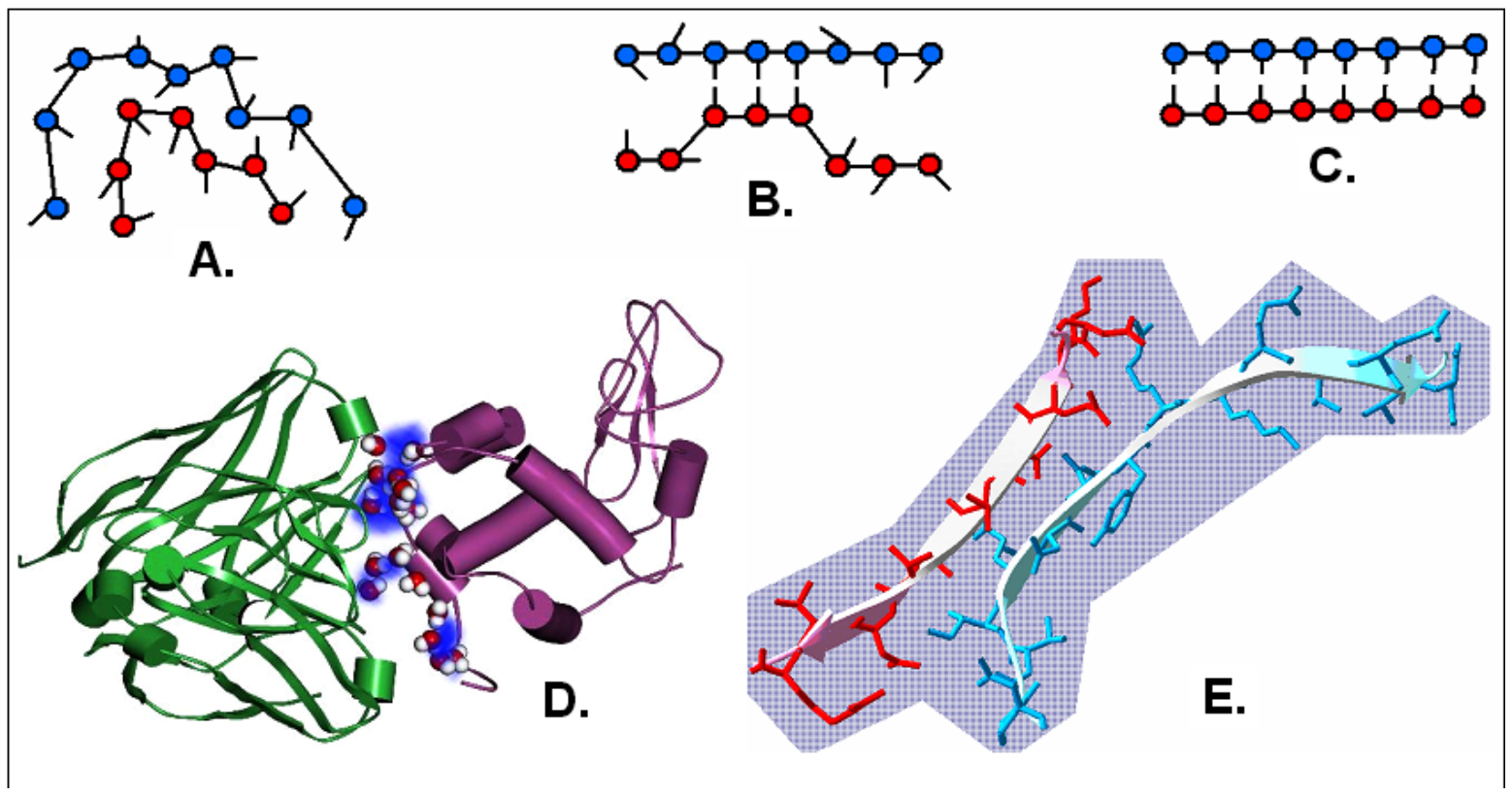

**Figure 7: Forms of peptide to peptide interactions.**

The specificity of interactions between two peptides might be explained in two ways. First, many amino acids collectively form larger configurations (protrusions and cavities, charge and hydropathy fields) which fit each other (A and D). Second, the physico-chemical properties (size, charge, and hydropathy) of individual amino acids fit each other like “lock and key” (C and E). There are even intermediate forms (B).

Is it really plausible that there is physicochemical compatibility between individual amino acids on the interacting surfaces? Bioinformatical studies clearly indicate that co-locating (interacting) amino acids are preferentially compatible (complementary) to each other regarding their size, charge and hydropathy [42].

This small-scale compatibility suggests that large, complex structure-forming is not an absolute structural requirement of specific protein interactions, but relatively strong and specific interactions might be formed already between a few (10-15) amino acid long oligo-peptides.

The present system is a unique combined *in silico* and *in vivo* method for identifying binding proteins that interact most effectively with reactive epitopes on a respective protein antigen. The system has widespread applications and is beneficial to biotechnology. It is useful, for example, in developing drugs and diagnostic kits for medical purposes. It has applications related to environmental health and public safety, including for example the detection of bacteria, viruses, toxins, etc. in air, water and food supplies.

The idea of Proteomic Code is not new and it was successfully implemented in numerous cases (for review see [1]). The recent “second generation concept” is a new significant improvement compared to the original concept. However it is clear that, as in the case of any new methods, a large number of data and evidence (synthetic peptides, western blots, SELDI, SPR etc) are needed to prove the general validity of this approach.

What about if the concept of Proteomic Code is wrong? The SHARP® design Method still remains a novel variant of combinatorial protein engineering, an approach proved to be successful for affinity peptide design and production.